\documentclass[utf8]{frontiersFPHY} 

\setcitestyle{square} 
\usepackage{lineno,microtype,subcaption}
\usepackage[onehalfspacing]{setspace}

\usepackage{listings}

\usepackage{url}
\usepackage[breaklinks]{hyperref}

\usepackage{breakurl}

\def\keyFont{\fontsize{8}{11}\helveticabold }
\def\firstAuthorLast{Dong {et~al.}} 
\def\Authors{Zhihua Dong\,$^{1}$, Heather Gray\,${^2}$, Charles Leggett\,$^{2, *}$, Meifeng Lin\,$^{1}$, Vincent R. Pascuzzi\,${^2}$ and Kwangmin Yu\,$^{1}$ }

\def\Address{
$^{1}$Brookhaven National Laboratory, Upton, NY, USA 

$^{2}$Lawrence Berkeley National Laboratory, Berkeley, CA, USA
}

\def\corrAuthor{Charles Leggett}
\def\corrEmail{cgleggett@lbl.gov}

\begin{document}
\onecolumn
\firstpage{1}

\title[Porting HEP Parameterized Calorimeter Simulation Code to GPUs]{Porting HEP Parameterized Calorimeter Simulation Code to GPUs} 

\author[\firstAuthorLast ]{\Authors} 
\address{} 
\correspondance{} 

\extraAuth{}
\maketitle

\begin{abstract}
The High Energy Physics (HEP) experiments, such as those at the Large Hadron Collider (LHC), traditionally consume large amounts of CPU cycles for detector simulations and data analysis, but rarely use compute accelerators such as GPUs. As the LHC is upgraded to allow for higher luminosity, resulting in much higher data rates, purely relying on CPUs may not provide enough computing power to support the simulation and data analysis needs. As a proof of concept, we investigate the feasibility of porting a HEP parameterized calorimeter simulation code to GPUs. We have chosen to use FastCaloSim, the ATLAS fast parametrized calorimeter simulation. 
While FastCaloSim is sufficiently fast such that it does not impose a bottleneck in detector simulations overall, significant speed-ups in the processing of large samples can be achieved from GPU parallelization at both the particle (intra-event) and event levels;
this is especially beneficial in conditions expected at the high-luminosity LHC, where extremely high per-event particle multiplicities will result from the many simultaneous proton-proton collisions.
We report our experience with porting FastCaloSim to NVIDIA GPUs using CUDA. A preliminary Kokkos implementation of FastCaloSim for portability to other parallel architectures is also described. 

\tiny
\keyFont{ \section{Keywords:} Large Hadron Collider, High Performance Computing, GPU, CUDA, Kokkos, Performance Portability, Particle Physics} 
\end{abstract}

\section{Introduction}
The ATLAS detector at the Large Hadron Collider (LHC) relies heavily on massive samples of simulated proton-proton collision events for detector modeling, upgrade studies and to deliver high-quality physics results.
However, standard simulations — using Geant4~\cite{Agostinelli:2002hh} for accurate and detailed modeling of physics processes and detector geometry — are extremely CPU-intensive, and its use for all Monte Carlo-based studies is impractical due to the formidable size of the LHC dataset.
In particular, simulating electromagnetic and hadronic shower developments in the ATLAS calorimeters is extremely CPU-intensive, comprising more than 90\% of the overall simulation time~\cite{ATLAS_Sim_Infrastructure}.

To reduce the significant processing time required to simulate the calorimeters,  ATLAS developed FastCaloSim~\cite{ATL-SOFT-PUB-2018-002,ATLAS_PUB_REPO}, a primarily C++ package that uses a simplified detector geometry and parameterizations of shower development initiated by particles traversing the calorimeter volumes.
The parameterizations are derived using single particles — electrons, photons and charged pions — with initial energy between 1~GeV and 4~TeV,  uniformly distributed in pseudorapidity, $\eta$.
During a simulation, the optimal parameterization is selected based on the particle type, energy deposit and location in the detector in order to best model the corresponding electromagnetic or hadronic shower.
The parameterizations are then used to create calorimeter \textit{hits}, which contain information about the location, and amount of energy deposited in the detector.
Thus, the more initial and secondary particles produced in a shower, the larger the number of hits.
Following the simulation, hits are then processed and used for event reconstruction.
Therefore, it is the number of hits, not necessarily the number of initial particles, that determines the length of time required for a particular simulation.

A parameterized simulation can reduce the CPU time by a factor of 10 to 25 — depending on the process being modeled — relative to that of fully Geant4-based simulations.
However, with respect to detector simulation workflows combining Geant4 for the Inner Detector and Muon Spectrometer with FastCaloSim, simulation of the calorimeters is no longer the
speed-limiting component.
Nevertheless, there are a number of advantages of porting FastCaloSim to accelerators, in particular, GPUs.
First, an intelligent port of the serial CPU code to GPU-friendly code permits parallelization over events (event-level parallelization) and also over particles within a single event (intra-event, or particle-/hit-level parallelization).
The ability to parallelize in these ways becomes exceedingly important when simulating not only a single proton-proton collision (the ``hard-scatter'' event) but when additional collisions within the same bunch-crossing are also simulated (``pile-up'' events).
For example, at the LHC there are on average 30~\cite{ATLAS_RUN_CONDITIONS} pile-up events for every one hard-scatter;
in the conditions of the High-Luminosity LHC (HL-LHC)~\cite{HLLHC_RUN_CONDITIONS}, there are expected to be more than 200 pile-up events for every hard-scatter. ATLAS foresees utilizing fast simulation for 60-90\% of its simulation workload for various HL-LHC scenarios~\cite{ATLAS-TDR}, making highly-parallelizable codes  of this nature extremely desirable. 
Lastly, an initial port of a set of codes such as those comprising FastCaloSim, that makes significant use of GPUs and demonstrates their computational ability, can be invaluable to the high-energy physics community in general.
It can serve as a proof of concept and motivation for other groups and collaborations to design and implement their own codes to execute on accelerator hardware, and prompt investigations into alternative portable solutions to exploit the wide range of modern architectures that are becoming increasingly pervasive.

\section{Test Data}\label{sec:test-data}
The inputs to FastCaloSim are tuples containing information about the incoming particle: the particle type and associated four-vector,
which consists of the particle's energy and momentum.
    Additional details -- \textit{e.g.}, calorimeter layer, cell and amount of energy deposited -- are also included to determine the correct parameterization as the particle traverses the detector.

For the studies presented here, several different inputs were used; single-particle samples -- electrons, photons and charged pions -- with various initial energies between 16~GeV and 4~TeV, and $\eta$ within the acceptance of the calorimeter; and a sample consisting of $t\bar{t}$ -- top-anti-top quark -- production at a center-of-mass proton-proton collision energy of 13~TeV.
The underlying event and pile-up (\textit{i.e.}, multiple proton-proton collisions in the same bunch-crossing) are not included in these simulations.
This is a configuration similar to that as was used in Reference~\cite{ATL-SOFT-PUB-2018-002}. 
Single-particle inputs contain only one primary initial particle that results in a relatively small number, $\mathcal{O}(1)$,
of secondaries being produced per event as the primary particle interacts with the calorimeter material.
In contrast, the $t\bar{t}$ simulation results in more realistic (complex) final states, with $\mathcal{O}(1000)$ secondaries produced per event.
Such large per-event particle multiplicities could potentially utilize more GPU compute -- provided they have sufficient energy to traverse, and not stop in, the calorimeter -- and can therefore represent a good benchmark for algorithm performance.

\section{Performance Profiling of the Unaltered FastCaloSim}
In the ATLAS production runs, FastCaloSim is part of the Athena software framework~\cite{ATLAS_PUB_REPO}. In order to make it easy to experiment with the GPU porting, we used the standalone version of FastCaloSim to reduce the number of dependencies that are typically needed in the Athena framework. The version of the standalone FastCaloSim package we used consists of roughly 40,000 lines of C/C++ code.

In order to determine where the computational hotspots are located in FastCaloSim, we ran performance profiling using the test data described in Sec.~\ref{sec:test-data} for electrons with an energy of 65.536~GeV and $\eta_\text{min}=0.2$. The test was run using a single core on Intel Xeon CPU E5-2695 v4 at 2.10 GHz. We generated call tree information using \texttt{valgrind --tool=callgrind}, and visualized the results with  \texttt{KCachegrind}. We found that more than 40\% of time was spent reading and unpacking the input ROOT data files, while the actual simulations consumed about 57\% of the total execution time. The focus of this work and subsequent discussions is on accelerating the simulation part, which is amenable to parallelization and  is contained in about $\mathcal{O}(1000)$ lines of code. The I/O overhead will limit the overall speedup we can achieve on the GPUs, which will also be discussed. 

\section{CUDA Implementation}
For porting to the GPU we focused on the function which, other than data unpacking, consumed the majority of the execution time: \texttt{TFCSLateralShapeParametrizationHitChain::Simulate\_hit()}  (see Listing \ref{hitsiml}). This function performs the hit simulation  in a calorimeter layer for a passing particle. In this function there is a loop over the number of hits which can range from 1 to 15,000  depending on particle type, energy, angle and calorimeter layer. For the input sample file of  electrons with  $E=65.536$ GeV and $\eta_\text{min}=0.2$, the number of hits typically range from 4000--6500.   Each hit  loop entry will decide into which calorimeter cell unit this simulated hit will deposit energy. The calculation is based on particle position, energy, angle and a random parametrization.  The hit simulation loop is only offloaded to the GPU when the number of hits is over 500 for a particular layer. Each GPU thread will simulate one hit.  If the number of hits is less than 500, or it is in a forward detector layer (layers 21-24)~\cite{ATLAS_Detector}, the calculation is left on the host. This is to ensure that the GPU can offer computational speedup compared to the CPU, which can only be achieved if there are a large enough number of hits (GPU threads) to keep the GPU busy. 

\begin{lstlisting}[language=C++, frame=single, caption=Hit simulation function, basicstyle=\small, label=hitsiml]
TFCSLateralShapeParametrizationHitChain::simulate( 
                                TFCSSimulationState& simulstate
                                const TFCSTruthState* truth,
                                const TFCSExtrapolationState* extrapol ) {																		
    int nhit = get_number_of_hits( simulstate, truth, extrapol );
    float Ehit = simulstate.E( calosample() ) / nhit;
    
    for ( int i = 0; i < nhit; ++i ) {  //  hit simulation loop
      TFCSLateralShapeParametrizationHitBase::Hit hit;
      hit.E() = Ehit;
      for ( TFCSLateralShapeParametrizationHitBase* hitsim : m_chain ) {
         hitsim->simulate_hit( hit, simulstate, truth, extrapol );
      }
    }
}
    
\end{lstlisting}

The hit simulation calculation requires detector geometry data and  parametrization tables. To offload  the calculation, this data first needs to be loaded onto the GPU. 
 
 For the  data used, the detector geometry included  187,652  calorimeter elements and 24 layers or samples, located in  78 regions.
 However, on the host side, the detector geometry data structures make heavy use of the C++ Standard Template Library (STL)  and pointers.  For example, the host region data structures include  \texttt{std::vector<std::vector<CaloDetDescrElements *>>}.  These kinds  of data structures are not supported by CUDA - in general, STL data structures will automatically re-allocate memory as their size changes, a feature not supported on GPUs due to their very limited memory resources. Furthermore, copying many small  chunks  of  non-contiguous  data generated with \texttt{new()} from CPU to GPU is  very inefficient.   We thus flattened the data, using arrays instead of pointers, and moved scattered  data together  before copying them to the GPU.
 After being copied to the GPU, all calorimeter detector element data as well as all region data were kept together as array of \texttt{structs}. The total size of the geometry data was about 20MB.
   
 Similarly, the parametrization tables, which are held in an \texttt{std::vector}, were transformed into a simple array before being copied to the GPU. Since the complete parametrization data is very large, only the needed tables were loaded onto the GPU during simulation. Furthermore, the same table is only loaded once, so if subsequent events require the same tables, they will not be loaded again. The total size of tables loaded to GPU depends on the type of simulations. Single particle simulations have a small footprint, while the data size for multi-particle simulations can be big. In our benchmark study for electrons with $E=65.536$~GeV, a simulation for 10,000 events only needs 1.2MB of table data loaded to the GPU,  while in the $t\bar{t}$ simulation with center-of-mass energy of 13~TeV, 500 multi-particle events need to have 1.1GB of tables loaded. 
 
 Hit simulation requires a large amount of random numbers. Each hit needs 3 random numbers.  The \texttt{cuRand}  library function was used to generate random numbers on the GPU.  However this will only produce statistically equivalent and not binary identical result when compared to CPU only calculations.  For code and physics validation, random numbers were also generated on the host, and copied to the GPU for calculations.  This technique was only used for validation, and not for performance benchmarks, as it is much slower.

 Other than the random number generator, there are 3  CUDA kernels in the code. The first, \texttt{clean}, resets the workspace. The number of threads is the same as the total number of detector cells.  The second or main kernel, \texttt{sim\_a}, does the hit simulation,  which calculates which cell gets the hit, and deposits energy into it. Each thread simulates one hit.  Atomic operations were chosen to accumulate the energy deposits, as the number of hits is relatively small. A block reduction method  was also tried, but did not demonstrate any performance gain, and introduced more kernels.  The last kernel,   \texttt{sim\_ct},   counts the hit cells and gathers the result for more efficient data transfer back to the host. The total number of threads is again the same as the total number of detector cells.
 The kernels are launched each time the code enters  the hit simulation loop, \textit{i.e.}, for each calorimeter sample layer a particle is passing through. After the third kernel executes, the results, an  array of   \texttt{struct \{long long  cell\_id;  float  E \} } are copied back to host.  For all 3 kernels, we used block size 256, \textit{i.e.}, 256 threads per block, in our benchmark. 
 Since the algorithmic processing is sequential in nature, the kernels are launched synchronously. Using asynchronous kernel launches and pre-loading data for the next event would have required large scale reconfiguration of the code, and was beyond the scope of this study.
 
To avoid many small kernel launches that would be required if the random numbers were generated on demand, GPU memory was pre-allocated, and a large number of random numbers were generated before hand, so that they would not need to be generated in every hit simulation loop. Regeneration of random numbers is only triggered when they are all consumed. In our benchmark test, we generate 200,000 random numbers each time. The infrequent generation of these random numbers places a very small burden on the GPU, and does not significantly affect performance.

We noticed that for some combinations of (particle type, energy, $\eta$),  the numbers of hits were very small. Since we only offloaded the simulation to the GPU when there were a minimum of 500 hits, most  of the calculation could still be done on the host.  To solve this problem, we explored the method of grouping hit simulations. We combined  different layers, particles, events together so that each GPU simulation  kernel launch can process more hits, with tasks of multiple entries to the simulation function as shown in Listing \ref{hitsiml}.  In order to combine these tasks, every time the hit simulation part is executed, instead of directly  doing the calculations on either the GPU or the CPU, we only collect the parameters needed for simulation, such as particle type, position, layer, number of hits, energy etc.  The  simulation calculation is delayed until enough hits have been accumulated. This information is then loaded onto the GPU and the simulation kernel is launched with more threads. Each thread can discover its own parameters needed for the hit it simulates and perform proper simulation. Our preliminary results in Sec.~\ref{sec:cuda_bench} show that this technique can lead to positive gains.  

\section{Performance Benchmarks}\label{sec:cuda_bench}

To perform the benchmarks, we used the GPU nodes on the Brookhaven National Lab Institutional Cluster (BNL-IC) and the NERSC Cori GPU cluster.  The BNL-IC node has 2 Intel\textsuperscript{\textregistered} Xeon\textsuperscript{\textregistered} CPU E5-2695 processors, each with 18 cores.  It  also  has 2 NVIDIA P100 GPUs.  Only one GPU was used when running a single instance of the code.
Two GPUs were used only when we tested multiple instances of the program running  with  \texttt{nvidia-cuda-mps-server}.

Cori GPU nodes  have two Intel\textsuperscript{\textregistered} Xeon\textsuperscript{\textregistered} Gold 6148 CPU and 8 NVIDIA V100 GPUs.  Our benchmark job used only one CPU core and one GPU.

The program starts by reading parametrization data files, unpacking them, and setting up the simulation chains. It then goes through an event loop to perform simulation for each event. The accumulated results  are saved  after event loop is completed.  Other than the event loop, the majority of the time was spent on unpacking data which doesn't change much with different  input sample files.

Our benchmarks collect time  spent in the event loop  and for the complete run by wrapping the code sections with calls to \texttt{std::chrono::system\_clock::now()}.

Table \ref{tab:allp} shows the result from running on the BNL-IC, comparing the original CPU  code and GPU accelerated code run time for a few different  particles and pseudorapidities with an energy of less than or equal to $E = 65.536$~GeV.
Multi-particle ($t\bar{t}$) input data  has 500 events while all others have 10,000 events.
For multi-particle data which simulate $t\bar{t}$ events with center-of-mass energy of 13~TeV, each event has about 600-800 particles with different energies and angles. 
All  timings were obtained  by running on the same node. Numbers were  averaged over ten repeated runs. Note that some of the runs had anomalously long run times, which are included in the averages. The fluctuations of the run times are reflected in the standard deviations shown in the parentheses. The longer run times can be attributed to I/O, which is needed in the event loop to read in some input files from the disk, and may be more susceptible to other activities in the system. Some examples of the I/O time variations are shown in the Appendix.

  \begin{table}[h]
\renewcommand{\arraystretch}{1.3}
\fontsize{8.2pt}{9pt}\selectfont{
\begin{center}
    \begin{tabular}{| r | c | c |c|c|c|c| c || c|c | c |c|c|}
    \hline
   &$E$&  & \multicolumn{3}{|c|}{Total Time [Seconds]} &\multicolumn{2}{|c|}{Speed Up}& \multicolumn{3}{|c|}{Event Loop time [Seconds]}&\multicolumn{2}{|c|}{Speed Up} \\
    \cline{4-13}
    \multicolumn{1}{|c|}{Particle} &  [GeV]& $\eta_\text{min}$ & CPU & GPU & GPU$_g$&GPU&GPU$_g$& CPU & GPU& GPU$_g$&GPU&GPU$_g$ \\
    \hline
    Electron         & 65.536 & 0.0  & 34.6(2) & 19.26(7) &17.5(2) & 1.80(1) & 1.97(2) & 18.30(9) &2.71(4) & 0.90(1) & 6.8(1) & 20.3(3) \\
    Electron          & 65.536 & 0.2 & 35.0(2) & 18.98(9) & 17.6(1) & 1.84(1) & 1.99(2) & 18.7(1) &2.37(5) &0.97(2) & 7.9(2) & 19.2(5) \\
    Electron        & 65.536 & 2.2 & 36.1(2) & 19.1(1) &17.8(1) & 1.89(1) & 2.02(1) & 19.7(1)  &2.56(3)&1.19(2) & 7.7(1) & 16.6(3)\\
    Electron        & 65.536 & 3.4 &18.0(1) &18.2(1)  & 18.1(1) & 0.985(8)& 0.990(9)& 1.60(4) &1.64(3)&1.54(2) & 0.97(3) & 1.04(3) \\
    \hline
     Photon         & 32.768 & 0.2 &26.2(4) &18.6(2) &18(3) & 1.41(3) & 1.4(2) &9.7(3)  &2.1(2) &0.8(2) & 4.7(6) & 12(3) \\
     Photon         & 65.536 & 0.2 &34.9(3) &18.9(2) &17.8(9) & 1.85(2) & 2.0(1)  &18.6(3)  &2.3(2) &1.0(2) & 8.0(7) & 19(4)\\
     \hline
     Pion         & 16.384 & 0.2 &19.02(7) &18.4(1) &17.5(1) & 1.033(9) & 1.089(7)&2.67(2)  &1.762(9) &0.89(2) & 1.52(1) & 2.99(6)\\
     Pion         & 32.768 & 0.2 &21.1(2) &18.83(9) &17.7(1) & 1.12(1) & 1.19(1) &4.57(5)  &2.17(3) &1.08(2) & 2.10(4) & 4.25(9) \\
     Pion         & 65.536 & 0.2 &24.1(1) &19.3(1) & 17.9(1) & 1.25(1) & 1.35(1) &7.68(6)  &2.70(5) &1.30(2) & 2.85(6) & 5.91(9) \\      
     \hline
     \hline
     M.P. &&& 53(1)& 46(1)&36.2(9)&1.14(4) & 1.46(4) &35.9(9)&29(1)&19.1(8) &1.22(6) & 1.88(9)\\
     \hline
    \end{tabular}
\end{center}
}
    \caption{ Execution time for the complete job and for only the event loop part of the job, for various input files. M.P. stands for multi-particle. 
    GPU$_g$ is for GPU group simulation code. Numbers in the parentheses are the significant digits of the standard deviations. For example, 34.6(2) means $34.6\pm0.2$, while 53(1) is equivalent to $53\pm 1$.}
    \label{tab:allp}
\end{table}

In Table \ref{tab:allp}, the CPU code event loop time roughly reflects the total number of hits.
It is important to note that the ATLAS forward calorimeters, with acceptance $3.1 < |\eta| < 4.9$, are not considered in this version of FastCaloSim.
As a result, the electron sample with $\eta_\text{min}=3.4$ shows comparable event loop times for CPU and GPU$_{(g)}$.
We can see that for the data file with more hits, such as electrons and photons with higher energy  and smaller $\eta$, both GPU and group simulation, GPU$_g$, can reduce total run time by 40-50\% compared to CPU code. For the event loop part of the run, GPU code can run faster than CPU code by a factor of 6--8 while group simulation can improve this by another factor of 2--3.  For  input data with fewer hits, such as Pions, the gain is less.  For  the multi-particle data we found that in the simulation, half of the layers only received 1 hit and many  layers have small numbers of hits, with the average hit count to be about 50. This is due to many low energy particles in each event, resulting in a low GPU simulation rate.  Furthermore, a lot more parametrization data need to be loaded onto  the  GPU  due to the variety of particles, energies and angles.   Hence in this case  the speedup on event loop is only a factor of 1.2 for GPU code and 1.9 for GPU$_g$ code.     

Table \ref{tab:nsc} shows the results of running on the NERSC Cori GPU node which are consistent with the BNL-IC results. Note that Cori has slightly faster CPUs and GPUs.

\begin{table}[h]
\renewcommand{\arraystretch}{1.3}
\begin{center}
    \begin{tabular}{| r | c | c || c | c| c || c |c|c||}
    \hline
    &$E$&& \multicolumn{3}{|c||}{Total Time [Seconds]} & \multicolumn{3}{c||}{Event Loop time [Seconds]} \\
    \cline{4-9}
    \multicolumn{1}{|c|}{particle} &[GeV]& $\eta_\text{min}$ & CPU & GPU & GPU$_g$& CPU & GPU & GPU$_g$ \\
    \hline
    Electron          & 65.536 & 0.2 & 34.1(4) &18.17(3)&  16.9(5) & 18.0(2) &2.02(2) &0.88(2)  \\
    \hline
    M.P.   &&& 51(2) &42.4(4)& 34.7(9) &  35(2) & 26.2(1) &18.1(1) \\
     \hline
    \end{tabular}
\end{center}
    \caption{ Execution time of complete run and event loop on Cori GPU node. M.P. stands for multi-particle.}
    \label{tab:nsc}
\end{table}

Simulations with higher energies at $2^{20}, 2^{21}$ and $2^{22}$ MeV ( or roughly 1~TeV, 2~TeV and 4~TeV) for fewer particles were also benchmarked  on the BNL-IC. Results are shown in Table \ref{tab:high}. Similar to previous benchmarks shown in Table~\ref{tab:allp}, some of these runs had large I/O time in the event loop, which are reflected in the standard deviations shown in the parentheses. The variations of the I/O time are shown in the Appendix for some representative runs. 
Comparing Table~\ref{tab:allp} and Table~\ref{tab:high}, we can clearly see that the GPU code speed-up for high-energy simulations is much better than for the lower energy ones. On average, speed-ups of up to a factor of 60 can be achieved for the event loop, and 4.6 for the total run. 
It should be noted that group simulation code does not gain as much over normal GPU code for electrons and photons compared to lower energy simulations.
In the case of higher energy particle simulations, such as with the 4~TeV electrons, performance of the group simulation can be worse than the regular one, due to the fact that each layer already has many simulated hits which make normal GPU code calculation already efficient, while the group simulation code needs to perform extra steps collecting simulation parameters.

\begin{table}[h]
\renewcommand{\arraystretch}{1.3}
\fontsize{8.75pt}{9pt}\selectfont{
\begin{center}
    \begin{tabular}{| r | c | c |c|c|c|c| c || c|c | c |c|c|}
    \hline
   &$E$&\# of & \multicolumn{3}{|c|}{Total Time [Seconds]} &\multicolumn{2}{|c|}{Speed Up}& \multicolumn{3}{|c|}{Event Loop time [Seconds]}&\multicolumn{2}{|c|}{Speed Up} \\
    \cline{4-13}
    \multicolumn{1}{|c|}{Particle} &  [TeV]& Events & CPU & GPU & GPU$_g$&GPU&GPU$_g$& CPU & GPU& GPU$_g$&GPU&GPU$_g$ \\
    \hline
Electron&	1&	3000    &81.4(4)&   18.02(5)&	17.6(2)&	4.51(2)&	4.62(4)&	65.1(3)&	1.58(2)&	1.1(1)&	41.2(6)&	59(6)\\
Electron&	2&	1000	&51.0(8)&	17.12(6)&	17.1(2)&	2.98(5)&	2.98(6)&	34.6(3)&	0.699(7)&	0.6(1)&	49.5(7)  &55(12)\\
Electron&	4&	900     &60.5(4)&	17.18(7)&	17.3(6)&	3.52(3)&	3.5(1) &	44.2(5)&	0.714(5)&	0.8(6)&	62.0(8) &	53(37)\\
Photon&	    1& 	3000    &81(1)&	18.2(3)&	17.5(4)&	4.4(1)&	4.6(1)&	63.8(4)&	1.6(2)&	1.1(3)&	40(6) &	60(18)\\
Photon&	    2&	1000	&50.4(4)&	17.3(2)&	17.1(4)&	2.92(3)&	2.94(8)&	34.1(4)&	0.8(2)&	0.6(4)&	44(9) &	56(33) \\
Photon&	    4& 	1000	&64.5(4)&	17.4(4)&	17.3(3)&	3.71(8)&	3.72(7)&	48.2(4)&	0.9(3)& 	0.8(3)&	53(17)&	57(21)\\
Pion&	    1& 3000&	37(1)&	18.47(7)&	17.5(1)&	2.03(8) &	2.13(8)&	21.2(13)&	1.99(6)&	1.08(5)&	10.6(7)&	20(2)\\
Pion&	    4& 1000&	30.8(2)& 	17.2(2)&	16.9(2)&	1.80(2)&	1.83(2)&	14.5(2)&	0.7(2)&	0.4(1)&	22(5)&	39(10)\\
\hline
    \end{tabular}
\end{center}
}
    \caption{ Execution time of complete run,  event loop  and speed up over CPU only code for high energy input files.
     All simulations are with $\eta_\text{min}= 0.2$.  Speed up is over CPU code. Numbers in the parentheses are the significant digits of the standard deviations. For example, 81.4(4) means $81.4\pm0.4$, while $81(1)$ represents $81\pm1$.}
    \label{tab:high}
\end{table}

In all cases  above, a single instance of the job will only use  a small portion of the available GPU resources.  Current high energy physics production calculations are performed  in the High Throughput Computing (HTC) environments  where many jobs are run on each multi-core node at the same time, usually with one thread or one process per core.  To emulate this environment, we started multiple concurrent instances of the same job, and shared the GPUs on the node between them by means of \texttt{nvidia-cuda-mps-server}~\cite{cuda-mps}.

\begin{figure}[h!]
\begin{center}
\includegraphics[width=0.8\textwidth]{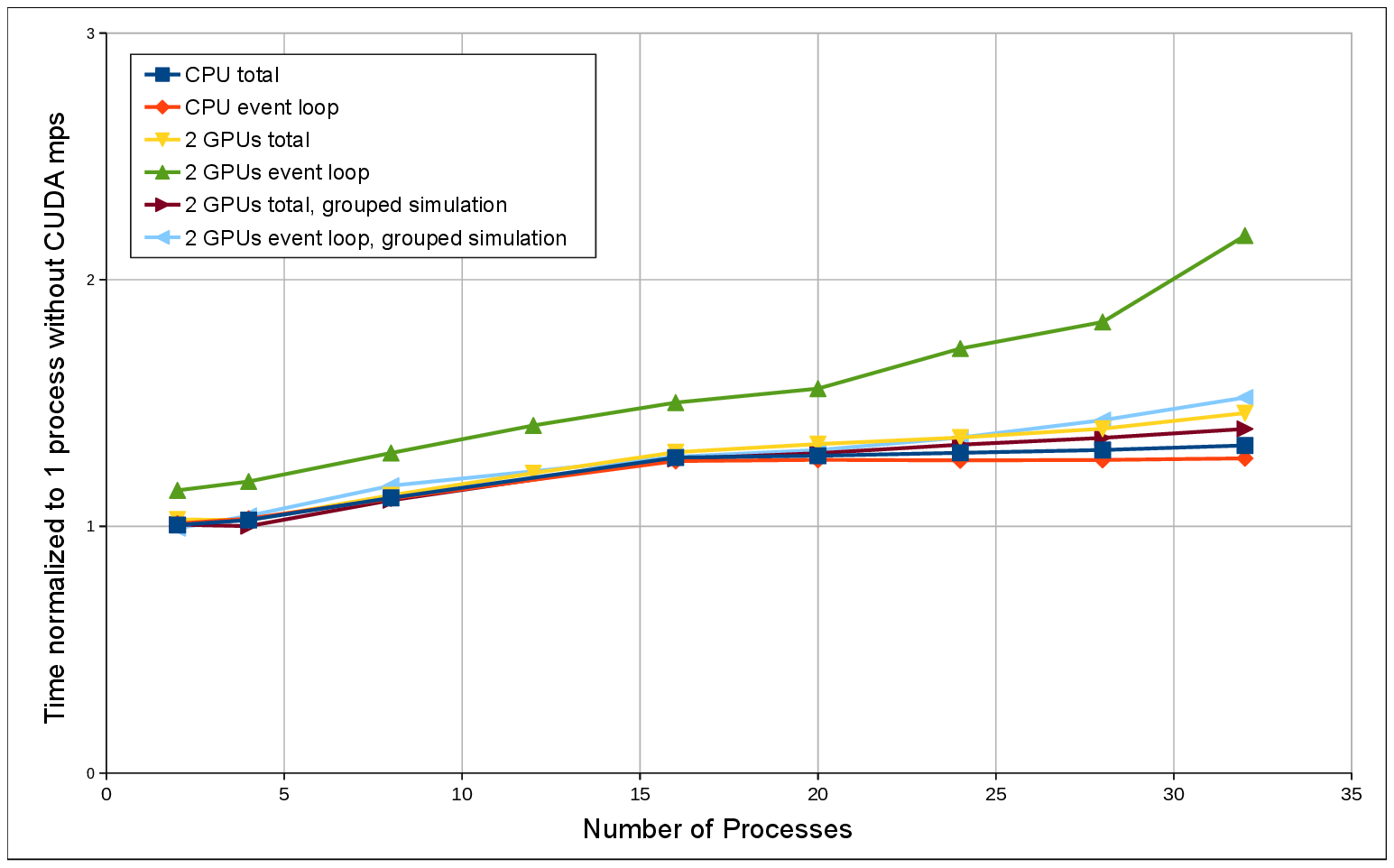}
\end{center}
\caption{ Normalized timing of multiple concurrent jobs running on the same node sharing GPUs with \texttt{nvidia-cuda-mps-server}}\label{fig:nproc}
\end{figure}

Figure \ref{fig:nproc} shows the result of running the benchmark on the same BNL-IC node, launching 2-32 processes of the program with CPU, GPU and group simulation codes.
The input sample data consisted of single-electrons with $E=65.536$~GeV, $\eta_\text{min}=0.2$ and 10,000 events. Since \texttt{nvidia-cuda-mps-server} for the NVIDIA P100 GPUs only supports a maximum 16 processes, we used 2 GPUs for a  maximum of 32 processes for our benchmark.  We can see that the CPU code slows down and the curve flattens as the number of concurrent jobs increases. This is due to core clock speed throttling from turbo boost when more cores are busy.  
The event loop time for per-event based simulation GPU code increased by a factor of more than 2 when the number of processes increased to 32, while the performance of grouped simulation codes is not significantly degraded as the number of concurrent jobs increase.  
While both codes have same amount of calculations, the per-event GPU code has significantly more small kernel launches and small data copy API calls than the grouped simulation. For many of them, the kernel launch latency overheads are greater than the actual kernel run times and data transfer times. The shift of the per-event GPU code (green triangles) indicates that the cuda-mps overhead is fairly significant compared to the GPU run time, while such overhead is negligible in the other cases where run times are significantly longer. The fact that we can run multiple concurrent processes without increasing much the execution time demonstrates that the computing units of GPUs are not fully utilized by any single job instance, even when shared among many jobs. However the overheads of the large number of small kernel launches and small data transfers start to saturate the cuda-mps managed GPUs when the number of processes reaches around 20 for the per-event based simulation. For that simulation we pinned the \texttt{nvidia-cuda-mps-server} processes as well as the simulation processes to the non-uniform memory access domains corresponding to  the GPU devices that they are associated with.

\section{Validation of the CUDA Implementation}
Since GPU ported code uses the \texttt{cuRAND} random number generator, which creates a different random number set from the CPU  simulation, we do not expect that the simulation results should be exactly comparable between the CPU and GPU. We should, however, be able to produce statistically consistent results, which we have checked and confirmed. 
In order to further validate the GPU code, we used the option of copying the same  set of CPU generated random numbers to the GPU, performing the simulation on the GPU, and comparing the results with the simulation performed solely on the CPU.  

We compared the results of around 62 million hits in the Electron 65.536~GeV run,  and found that  in the GPU simulation, there were two hits whose calculations ended up in different calorimeter cells compared to the CPU calculation. Further investigation showed that this was  due to floating point rounding error differences between CPU and GPU. The two mismatches can be eliminated by using double precision (\texttt{double}) variables for certain calculations instead of single precision \texttt{float}.  We can thus conclude that the GPU ported code is correctly performing the same calculations as the original CPU code.

\section{Portable Implementation with Kokkos}

Kokkos~\cite{kokkos} is a  C++ library, developed at Sandia National Laboratory, with a rich history 
and a strong user base. It is a
portable, performant, single source shared-memory programming model that abstracts away the hardware, allowing developers to write algorithms once and run on
any supported backend architectures. Currently supported backends are serial processing on a traditional
CPU, parallel processing on multi and many-core CPUs via pThreads and OpenMP, and NVIDIA and
AMD GPUs via CUDA and HIP. An Intel GPU backend is currently under development as well. 
A common API that hides the architecture-specific implementation
details greatly reduces the intimate device knowledge that would otherwise be needed by developers. 
At compile time, Kokkos permits the selection of one host serial 
backend (\textit{e.g.}, Intel Haswell architecture), one
host parallel backend (\textit{e.g.}, OpenMP), and one device parallel backend (\textit{e.g.}, NVIDIA V100). If different hardware
backends are desired, both the user code and Kokkos runtime library must be recompiled.

Kokkos has a number of different parallel abstractions available,
such as \texttt{parallel\_for}, \texttt{parallel\_reduce}, and \texttt{parallel\_scan}. It also provides
utilities such as random number generators, and supports
atomic operations. Data access is accomplished via multi-dimensional structures that are templated in the
type of the data called
\texttt{Kokkos::Views}. The sizes of the structures can be specified at either run or compile time.
Kokkos' usage of C++ language features in kernels are only limited by the compilers necessitated by
the selected backends -- \textit{e.g.}, if the CUDA backend is chosen for NVIDIA hardware -- and the available version
of the nvcc compiler is 10.2, then \texttt{C++14} is the latest standard that can be used in a kernel.

\subsection{Build Infrastructure}

FastCaloSim uses CMake as its build tool. Modifying the CMake configuration files to pickup the Kokkos
installation was relatively simple, as Kokkos provides the infrastructure support for CMake, with
\texttt{find\_package(Kokkos)}. 
Configuring Kokkos was more complicated. There are three build parameters of Kokkos that must be
selected when installing the library, which can affect how the FastCaloSim code is organized:
\begin{itemize}
    \item The hardware backends (host serial, host parallel, device parallel)
    \item The use of shared libraries
    \item The use of relocatable device code
\end{itemize}
For testing purposes, the choice of the hardware backends is not problematic, as the developer knows
where the code will run. It is more of an issue in a production environment where binaries might be
distributed to a broad range of hardware architectures, requiring multiple different configurations to be
built. The choice of using shared vs static libraries is usually dependent on what already exists in the
code base that will be using Kokkos. In this case, FastCaloSim uses shared libraries, and modifying
it to use static libraries would have made re-integration to the original code base much more 
challenging, so we decided to stay with shared libraries. Unfortunately, enabling shared libraries and
relocatable device code are mutually exclusive in the Kokkos build. Disabling relocatable device
code has significant implications for the structure of the kernel sources - all symbols in a kernel
must be visible and resolvable within a single compilation unit.

In order to keep all source files as synchronized as possible, we chose to keep the serial, CUDA and 
Kokkos codes in the same git branch, selecting which feature to build at CMake configuration time via
command line switches, such as \texttt{-DENABLE\_GPU=on} and \texttt{-DUSE\_KOKKOS=on}.

\subsection{Porting CUDA Kernels to Kokkos}

If the CUDA backend of Kokkos is selected as the device parallel
backend, CUDA and Kokkos code can freely interoperate, with Kokkos kernels having the ability to call
CUDA, and access data allocated via \texttt{cudaMalloc}. This feature allows for 
the incremental migration of CUDA kernels to Kokkos, simplifying testing and reproducibility. One 
by one, we rewrote the CUDA kernels as Kokkos lambda expressions and inlined functions,
to be executed as \texttt{parallel\_for} 
loops, but maintaining the CUDA memory allocations, and accessing the bare device pointers in
the kernels. At each step, we validated the outputs to make sure that the results were consistent with the original CUDA code.

Once all kernels had been migrated to Kokkos, we tackled the data structures. Most are simple scalars,
1D and 2D arrays. These were easy to translate to \texttt{Kokkos::Views}. A couple of data structures,
however, are jagged 2D arrays, \textit{i.e.}, arrays of arrays where the sizes of the inner arrays are not all the
same. While it is technically possible to create \texttt{Kokkos::View< Kokkos::View< T >>}, this is not
advised by the Kokkos developers, and requires a significant amount of extra boiler-plate code to create
and access. Instead we chose to flatten the jagged arrays, either by adding a new array which indexed
the starting locations of the inner arrays in a 1D structure when there was significant variation in the
sizes of the inner arrays, or by padding to the largest size of the inner arrays, when they were close
in size.

\subsection{Source File Organization}
Since Kokkos' use of relocatable device code was disabled, we had to structure the source files
so that all symbols were visible within a single compilation unit. This was accomplished by 
creating a wrapper file that merely included the various individual kernel source files. Since
the majority of the kernel device functions were identical for the Kokkos and CUDA implementations,
in order to share as much of the code as possible, the shared code was in a single file, with 
CUDA and Kokkos specific features triggered via preprocessor macros (see Listing \ref{devsel}).

\begin{lstlisting}[language=C++, frame=single, caption=Language functionality selection, basicstyle=\small, label=devsel]
#ifdef USE_KOKKOS
#  include <Kokkos_Core.hpp>
#  include <Kokkos_Random.hpp>
#  define __DEVICE__ KOKKOS_INLINE_FUNCTION
#else
#  define __DEVICE__ __device__
#endif

__DEVICE__ void HitCellMapping_d( Hit& hit, Chain0_Args args ) {
  long long cellele = getDDE( args.geo, args.cs, hit.eta(), hit.phi() );

#ifdef USE_KOKKOS
  Kokkos::View<float*> cellE_v( args.cells_energy, args.ncells );
  Kokkos::atomic_fetch_add( &cellE_v( cellele ), hit.E() );
#else
  atomicAdd( &args.cells_energy[cellele], hit.E() );
#endif
   ...
}
\end{lstlisting}

Implementations that were very different between the Kokkos and CUDA versions were kept in
separate files, and included by wrappers appropriately. 

Another complication is that Kokkos will compile source files differently depending on the 
suffix of the file name. If the suffix is \texttt{.cu}, the NVIDIA compiler \texttt{nvcc} will be 
used for the compilation with a specific set of command line switches that is different than if
the same file had a \texttt{.cpp} or \texttt{.cxx} extension. Compilation errors will occur
if Kokkos core headers are exposed to \texttt{nvcc} in \texttt{.cu} files, which required 
some careful organization
of file names and wrappers.

\subsection{Random Numbers}
A large number of random numbers are required during the execution of FastCaloSim. They
are used in code that is executed on the device, and are not needed on the host. Thus it would
make sense to generate them on the device, to avoid large memory transfers. Both CUDA and
Kokkos provide libraries to generate high quality random numbers. However, in order to verify
that different implementations produce the same results, it is very useful to be able to always
use the same set of random numbers. We thus provided the ability to generate random numbers
either on the host, and transferred to the device as needed, or directly on the device using
either the CUDA or Kokkos RNGs.

\subsection{Performance Studies}

Once we had fully migrated all kernels and data structures to Kokkos, we were able to test the
performance of the code on multiple different parallel devices. The hardware platform was
a dual Intel Xeon Gold 5220 CPU with 18 physical cores per CPU, for a total of 72 hyperthreaded
cores. The host OS was Centos~7, and packages were compiled with GCC 8.2. Also installed 
on the server were a NVIDIA RTX 2080 Super GPU, and an AMD Vega 56 GPU. We used Kokkos version 3.1
for testing purposes, which was configured to target the CPU with \texttt{Kokkos\_ARCH\_SKX}, and the GPUs
with \texttt{Kokkos\_ARCH\_TURING75} and \texttt{Kokkos\_ARCH\_VEGA900}. The installed version of CUDA
was 10.2.89, and of the ROCm libraries was 3.5.1. We also converted the CUDA code directly to HIP using
\texttt{hipify-perl}. For the sake of brevity, when we refer to HIP
in the subsequent sections, we mean the hcc backend of HIP that targets AMD devices, as opposed
to the nvcc backend that targets NVIDIA devices.

In Table \ref{tab:kokkos_perf} we show the relative execution times of various kernels with
different Kokkos backends, using the
CUDA implementation as a baseline, and reporting times as a factor slower than that of
the CUDA implementation.

The aspects that we measured were the time it takes to re-initialize a large array of approximately
200,000 floats (\texttt{clean})
at the start of each event, the time to execute the main kernel (\texttt{sim\_a}), the time to execute the
reduction kernel (\texttt{sim\_ct}), the time it takes to copy the results back from the device to the host 
(\texttt{copy d$\rightarrow$h}), and finally the time for the full event loop of 10,000 events. This was
done by placing synchronization barriers around the kernels, and using host based timers to
measure the execution times. The best
performance using the pThread and OpenMP backends were achieved using 15 threads.

\begin{table}[h]
\renewcommand{\arraystretch}{1.3}
\begin{center}
    \begin{tabular}{| r | c | c | c | c | c || c || c | c |}
    \hline
    & \multicolumn{5}{|c||}{Kokkos (relative)} & & CUDA & CPU \\
    \cline{2-7}
    \multicolumn{1}{|c|}{kernel} & CUDA & HIP$^1$ & Serial & pThread & OpenMP & HIP$^2$ & ($\mu$s) & ($\mu$s)\\
    \hline
    clean           & 2.83 & 20.1 & 7.67 & 1.61 & 2.26 & 5.02 & 14.0 & \\
    sim\_a          & 1.17 & 37.9 & 36.9 & 11.2 & 11.5 & 37.0 & 49.2 & \\
    sim\_ct         & 1.35 & 8.93 & 12.9 & 1.66 & 1.99 & 3.99 & 15.2 & \\
    copy d$\rightarrow$h & 1.30 & 3.92 & 0.02 & 0.06 & 0.04 & 2.98 & 25.3 & \\
    event loop      & 1.16 & 11.9 & 7.26 & 2.61 & 2.79 & 6.83 & 321 & 1726\\
    \hline \hline
    event loop speedup vs. CPU & 4.63 & 0.45 & 0.74 & 2.06 & 1.93 & 0.79 & 5.38 & 1 \\
    \hline
    \end{tabular}
\end{center}
    \caption{Relative kernel average execution times with various Kokkos backends, as a factor of the pure CUDA implementation execution time. 
    HIP\textsuperscript{1} is for Kokkos with the HIP backend, whereas HIP\textsuperscript{2} is a direct port from CUDA to HIP. The last two columns show the absolute kernel runtimes for the original CUDA code, and for the native CPU implementation, averaged over 10,000 events. The last row shows the relative speedup of the event loop versus the native CPU implementation.}
    \label{tab:kokkos_perf}
\end{table}

We also directly measured the performance of the kernels using \texttt{nvprof} and \texttt{rocprof}, as
show in Table \ref{tab:gpu_perf}.

\begin{table}[h]
\renewcommand{\arraystretch}{1.3}
\begin{center}
    \begin{tabular}{| r | c | c || c |}
    \hline
    & \multicolumn{2}{|c||}{Kokkos} & \\
    \cline{2-4} 
    \multicolumn{1}{|c|}{kernel} & CUDA & HIP$^1$ & HIP$^2$ \\
    \hline
    clean           & 1.60 & 4.07 & 2.62 \\
    sim\_a          & 0.93 & 43.6 & 44.3 \\
    sim\_ct         & 1.43 & 3.58 & 2.41 \\
    copy d$\rightarrow$h & 1.02 & 1.65 & 1.32 \\
    \hline
    \end{tabular}
\end{center}
    \caption{Relative kernel average execution times with CUDA and HIP\textsuperscript{1} Kokkos backends, and a direct HIP\textsuperscript{2} port, as a factor of the pure CUDA implementation execution time, as measured by GPU profiling tools \texttt{nvprof} and \texttt{rocprof}.}
    \label{tab:gpu_perf}
\end{table}

From these measurements, we can extract several important conclusions. First, using Kokkos to 
initialize memory on the device does impose some significant penalties as compared to a simple
CUDA operation. This is due to extra steps required to create the host views before the
device memory is reset with a \texttt{Kokkos::deep\_copy}. In general though, while we do see some overhead from using Views and
some small penalties from launching Kokkos kernels, the
performance of the CUDA backend of Kokkos is reasonably close to that of the pure CUDA implementation.
As the kernels get smaller and simpler, \textit{e.g.}, comparing the performance of \texttt{sim\_ct} to that 
of \texttt{sim\_a}, the Kokkos overhead become more noticeable.
Furthermore, the performance of the serial backend of Kokkos performs considerably worse than the native
CPU implementation (about 35\% slower). As well, using the OpenMP or pThread backend with 15 threads (where
the greatest speedup was achieved) would only be cost
effective if there was no available work for the other CPUs on the host, otherwise throughput was about a factor of 7.5
worse than an equivalent number of single threaded (native CPU) jobs.

Second, the performance of AMD GPU is noticeably worse than that of the NVIDIA GPU. While it is
hard to do a direct comparison as the architectures are very different, the clock speeds and
number of computation units are within 25\% of each other, yet we see timing differences in
some cases of one and a half orders of magnitudes. Further tests of simple kernels seem to 
suggest that the kernel launch latencies on the AMD GPU are significantly higher than that
of the NVIDIA GPUs (see Table \ref{tab:launch_latency}). The clock frequency of the CPU on the
host also has a significant impact on the launch latency.

\begin{table}[h]
\renewcommand{\arraystretch}{1.4}
\begin{center}
    \begin{tabular}{| c | c || c | c || c |}
    \hline
    & &\multicolumn{2}{|c||}{Kokkos} & \\
    \cline{3-4}
     CPU freq & CUDA & CUDA & HIP & HIP$^2$ \\
     \hline
     2200 MHz & 5.6 & 9.4 & 152 & 88 \\
     3700 MHz& 3.4 & 5.3 & 60 & 30 \\
    \hline
    \end{tabular}
\end{center}
    \caption{Kernel launch latencies in microseconds as a function of CPU clock frequency, measured by timing the execution of 10,000 empty kernels.}
    \label{tab:launch_latency}
\end{table}

It is also worth noting that when the job was pinned to a single CPU core, the timings for the CUDA and Kokkos/CUDA did not change appreciably, however the time to execute the full event loop when utilizing the HIP backend of Kokkos increased by 30\%, suggesting that scheduling and launching kernels
on AMD devices places significant burden on the CPU.

\subsection{Porting vs Rewriting}

The method that we chose to translate the CUDA code to Kokkos was obviously non-optimal, and in
some ways
deliberately so. Part of the exercise was to understand the issues involved in porting a
project which may have a large amount of existing code from serial C++ to CUDA, and from CUDA to Kokkos, 
while attempting
to leave existing serial C++ and CUDA code unmodified as much as possible. Given the
relatively small size of the FastCaloSim code base, it would have been easier to completely rewrite
the entire code base, refactoring to optimize for Kokkos data structures and kernels. This would
no doubt have improved the performance of the Kokkos version. However, ultimate performance is
not the end goal of our study, but rather performant portability, reproducibility, and the burden
of code maintenance are more important metrics. With a large code base, it is very important
to make sure that it is easy to maintain compatibility between all implementation flavours,
and also to ease migration of code from serial C++$\rightarrow$CUDA$\rightarrow$Portability Layer.

\section{Conclusions}
We have successfully ported the ATLAS fast calorimeter simulation code, FastCaloSim, to use GPUs. The code was first ported to NVIDIA GPUs using the proprietary CUDA API. As with many HTC applications, the compute-to-I/O ratio of FastCaloSim is not very large, limiting the potential gains of using GPUs. Several techniques have been incorporated in our implementations to increase the speed and efficiency of the computations on the GPUs, as well as the GPU's overall utilization. The performance gains on the GPUs are greatly dependant on the types of simulations (particle type, energy, angle etc.). For the test cases we ran, we observed overall speed-ups by as much as a factor of five. But if we compare only the timing of the simulation part, the speedups on the GPUs were much more significant, as much as 60 times relative to executing with a single thread on a CPU core. The GPU was found to be underutilized, and a single GPU may be shared among many concurrent CPU jobs of FastCaloSim via \texttt{nvidia\_cuda\_mps\_server} with very little loss of performance.

We have also shown a preliminary Kokkos implementation that supports running on multiple architectures, including multicore CPUs, NVIDIA GPUs and AMD GPUs. Kokkos has proven to be a powerful tool for performant portability on a number of different architectures. While there is some overhead with utilizing its data structures as compared with raw memory allocation, these are relatively small and offset by the gains in portability. Direct comparisons of kernel execution time running on both NVIDIA and AMD GPUs show reasonably similar measurements between Kokkos and device specific CUDA/HIP implementations. Since the Kokkos implementation was designed as a port from CUDA, and not rewritten from the ground up, it is very likely that further optimization of the Kokkos code would be possible. The ability to run on both NVIDIA and AMD GPUs, as well as multicore CPUs via OpenMP or pThreads, with minimal changes to user code, vastly outweighs any small performance losses.

\section*{Acknowledgments}
This work was supported by the DOE HEP Center for Computational Excellence at Brookhaven National Laboratory, and Lawrence Berkeley National Laboratory under B\&R KA2401045.
 This work was also done as part of the offline software research and development programme of the ATLAS Collaboration, and we thank the collaboration for its support and cooperation.
 This research used resources of the National Energy Research Scientific Computing Center (NERSC), a U.S. Department of Energy Office of Science User Facility located at Lawrence Berkeley National Laboratory, operated under Contract No. DE-AC02-05CH11231.
 We gratefully acknowledge the use of the computing resources provided by Brookhaven National Laboratory. 
 A git repository for the source code is available~\cite{FCS_PUB_REPO}.

\clearpage

\appendix

\section*{Appendix: Breakdown of I/O and Compute Time in Event Loop}
Here we show some representative examples of the detailed breakdown of the I/O time and compute time inside the event loop for the three implementations, original CPU (denoted as CPU), CUDA code (denoted as GPU) and CUDA group simulation code (denoted as GPU$_g$). Table~\ref{tab:bkdn-1} shows the detailed timing breakdown for I/O and compute time during the event loop for the simulation of Photons with $E=32.768$ GeV and $\eta_{\rm min}=0.2$, corresponding to the 5th row in Table~\ref{tab:allp}. Table~\ref{tab:bkdn-2} shows the detailed timing breakdown for the simulation of Electrons with $E\approx4$ TeV and $\eta_{\rm min}=0.2$, corresponding to the 3rd row in Table~\ref{tab:high}. We can see that in both Table~\ref{tab:bkdn-1} and Table~\ref{tab:bkdn-2}, the compute time is relatively stable for different runs for all three implementations, while the I/O time can have some large spikes, most of which appear at the first few runs. There was, however, a large spike at Run 3 of the GPU$_g$ test in Table~\ref{tab:bkdn-2}, which may be due to other activities in the system. This contributes to the large standard deviations of the corresponding numbers in Table~\ref{tab:high}. 

  \begin{table}[h]
\renewcommand{\arraystretch}{1.3}
\fontsize{10pt}{10pt}\selectfont{

\begin{center}
    \begin{tabular}{| r | c | c| c || c |c|c||c|c|c||}
    \hline
     & \multicolumn{3}{|c||}{CPU} & \multicolumn{3}{c||}{GPU} & \multicolumn{3}{c||}{GPU$_g$} \\
    \cline{2-10}
    \multicolumn{1}{|c|}{Run} & I/O [s] & Comp. [s] & Tot. [s] & I/O [s] & Comp. [s] & Tot. [s] & I/O [s] & Comp. [s] &  Tot. [s]  \\
    \hline
   1        & 0.92 &9.57 & 10.49 &0.88 &1.86 &2.74 & 0.72 & 0.65 & 1.37 \\
   2        & 0.11 &9.46 & 9.57 &0.11 &1.93 &2.04 & 0.09 &0.64 & 0.73\\
   3        & 0.11 &9.59 & 9.70 &0.11 &1.84 &1.95 & 0.08 &0.68 &0.76 \\
   4        & 0.10 &9.47 & 9.57 &0.11 &1.87 &1.98 & 0.08 &0.67 &0.75 \\
   5        & 0.11 &9.60 & 9.71 &0.11 &1.88 &1.99 & 0.08 &0.63 &0.71 \\
   6        & 0.11 &9.54 & 9.65 &0.12 &1.89 &2.01 & 0.08 &0.64 &0.72 \\
   7        & 0.10 &9.56 & 9.66 &0.12 &1.89 &2.01 & 0.08 &0.65 &0.73 \\
   8        & 0.11 &9.49 & 9.60 &0.11 &1.86 &1.97 & 0.08 &0.68 &0.76 \\
   9        & 0.10 &9.50 & 9.60 &0.11 &1.88 &1.99 & 0.08 &0.63 &0.71 \\
   10       & 0.11 &9.33 & 9.54 &0.11 &1.87 &1.98 & 0.08 &0.67 & 0.75 \\
   \hline
   Avg.(stdev) & 0.2(3) &9.52(6) &9.7(3) &0.2(2) &1.88(2) &2.1(2) &0.1(2) &0.65(2) & 0.8(2) \\
     \hline

    \end{tabular}
\end{center}
}
    \caption{Breakdown of the Event Loop time (Tot.) into I/O time (I/O) and compute time (Comp.) for the simulation of Photons with $E=32.768$ GeV and $\eta_{\rm min}=0.2$, corresponding to the 5th row in Table~\ref{tab:allp}.}
    \label{tab:bkdn-1}
\end{table}

 \begin{table}[h]
\renewcommand{\arraystretch}{1.3}
\fontsize{10pt}{10pt}\selectfont{

\begin{center}
    \begin{tabular}{| r | c | c| c || c |c|c||c|c|c||}
    \hline
     & \multicolumn{3}{|c||}{CPU} & \multicolumn{3}{c||}{GPU} & \multicolumn{3}{c||}{GPU$_g$} \\
    \cline{2-10}
    \multicolumn{1}{|c|}{Run} & I/O [s] & Comp. [s] & Tot. [s] & I/O [s] & Comp. [s] & Tot. [s] & I/O [s] & Comp. [s] &  Tot. [s]  \\
    \hline
   1        & 0.69 &43.91 & 44.60 &0.13 &0.60 &0.72 & 0.29 &0.49 &0.79 \\
   2        & 0.33 &43.54 & 43.87 &0.13 &0.58 &0.71 & 0.12 &0.49 &0.61\\
   3        & 0.30 &44.48 & 44.78 &0.12 &0.59 &0.71 & 2.06 &0.48 &2.54 \\
   4        & 0.14 &43.61 & 43.75 &0.13 &0.59 &0.72 & 0.22 &0.49 &0.71 \\
   5        & 0.13 &43.50 & 43.63 &0.13 &0.58 &0.71 & 0.13 &0.48 &0.61 \\
   6        & 0.13 &44.19 & 44.32 &0.12 &0.59 &0.72 & 0.13 &0.48 &0.61 \\
   7        & 0.14 &44.77 & 44.90 &0.12 &0.60 &0.71 & 0.13 &0.49 &0.61 \\
   8        & 0.14 &43.85 & 43.99 &0.13 &0.58 &0.71 & 0.13 &0.55 &0.68 \\
   9        & 0.13 &43.83 & 43.96 &0.13 &0.59 &0.72 & 0.13 &0.49 &0.62 \\
   10       & 0.14 &44.39 & 44.52 &0.12 &0.59 &0.71 & 0.16 &0.48 &0.64 \\
   \hline
   Avg.(stdev) & 0.2(2) &44.0(4) &44.2(5) &0.124(4) &0.590(6) &0.714(5) &0.3(6) &0.49(2) & 0.8(6) \\
     \hline

    \end{tabular}
\end{center}
}
    \caption{Breakdown of the Event Loop time (Tot.) into I/O time (I/O) and compute time (Comp.) for the simulation of Electrons with $E\approx4$ TeV and $\eta_{\rm min}=0.2$, corresponding to the 3rd row in Table~\ref{tab:high}.}
    \label{tab:bkdn-2}
\end{table}


\begin{thebibliography}{11}
\expandafter\ifx\csname natexlab\endcsname\relax\def\natexlab#1{#1}\fi
\expandafter\ifx\csname urlstyle\endcsname\relax
  \expandafter\ifx\csname doi\endcsname\relax
  \def\doi#1{doi:\discretionary{}{}{}#1}\fi \else
  \expandafter\ifx\csname doi\endcsname\relax
  \def\doi{doi:\discretionary{}{}{}\begingroup \urlstyle{rm}\Url}\fi \fi
\expandafter\ifx\csname selectlanguage\endcsname\relax
  \def\selectlanguage#1{}\fi

\bibitem[{Agostinelli et~al.(2003)}]{Agostinelli:2002hh}
Agostinelli S, et~al.
\newblock {GEANT4--a simulation toolkit}.
\newblock {\em Nucl. Instrum. Meth. A\/} {\bf 506} (2003) 250--303.
\newblock \doi{10.1016/S0168-9002(03)01368-8}.

\bibitem[{{The ATLAS Collaboration}(2010)}]{ATLAS_Sim_Infrastructure}
{The ATLAS Collaboration}.
\newblock {The ATLAS Simulation Infrastructure}.
\newblock {\em The European Physical Journal C\/} {\bf 70} (2010) 823-874.
\newblock \doi{10.1140/epjc/s10052-010-1429-9}.

\bibitem[{{The ATLAS Collaboration}(2018)}]{ATL-SOFT-PUB-2018-002}
{The ATLAS Collaboration}.
\newblock {The new Fast Calorimeter Simulation in ATLAS}.
\newblock Tech. Rep. ATL-SOFT-PUB-2018-002, CERN, Geneva (2018).

\bibitem[{ATL(2020)}]{ATLAS_PUB_REPO}
{ATLAS Software Repository}  (2020).
\newblock \url{https://gitlab.cern.ch/atlas/athena}.

\bibitem[{{The ATLAS Collaboration}(2020)}]{ATLAS_RUN_CONDITIONS}
{The ATLAS Collaboration}.
\newblock {ATLAS} data quality operations and performance for
  2015{\textendash}2018 data-taking.
\newblock {\em Journal of Instrumentation\/} {\bf 15} (2020) P04003--P04003.
\newblock \doi{10.1088/1748-0221/15/04/p04003}.

\bibitem[{Apollinari et~al.(2017)Apollinari, BM-CM-)jar~Alonso, BrM-CM-<ning, Fessia,
  Lamont, Rossi et~al.}]{HLLHC_RUN_CONDITIONS}
Apollinari G, BM-CM-)jar~Alonso I, BrM-CM-<ning O, Fessia P, Lamont M, Rossi L, et~al.
\newblock {High-Luminosity Large Hadron Collider (HL-LHC): Technical Design
  Report V. 0.1}.
\newblock Tech. rep., Geneva (2017).
\newblock \doi{10.23731/CYRM-2017-004}.

\bibitem[{Calafiura et~al.(2020)Calafiura, Catmore, Costanzo, and
  Di~Girolamo}]{ATLAS-TDR}
Calafiura P, Catmore J, Costanzo D, Di~Girolamo A.
\newblock {ATLAS HL-LHC Computing Conceptual Design Report}.
\newblock Tech. Rep. CERN-LHCC-2020-015. LHCC-G-178, CERN, Geneva (2020).
\newblock \url{https://cds.cern.ch/record/2729668}.

\bibitem[{{The ATLAS Collaboration}(2008)}]{ATLAS_Detector}
{The ATLAS Collaboration}.
\newblock The {ATLAS} experiment at the {CERN} large hadron collider.
\newblock {\em Journal of Instrumentation\/} {\bf 3} (2008) S08003--S08003.
\newblock \doi{10.1088/1748-0221/3/08/s08003}.

\bibitem[{cud(2020)}]{cuda-mps}
{NVIDIA multi-process service}  (2020).
\newblock
  \url{https://docs.nvidia.com/deploy/pdf/CUDA\_Multi\_Process\_Service\_Overview.pdf}.

\bibitem[{Carter et~al.(2014)Carter, Trott, and Sunderland}]{kokkos}
Carter H, Trott C, Sunderland D.
\newblock {Kokkos: Enabling manycore performance portability through
  polymorphic memory access patterns}.
\newblock {\em Journal of Parallel and Distributed Computing\/} {\bf 74} (2014)
  3202--3216.
\newblock \doi{10.1016/j.jpdc.2014.07.003}.
\newblock \url{https://github.com/kokkos}.

\bibitem[{FCS(2020)}]{FCS_PUB_REPO}
{FastCaloSim Software Repository}  (2020).
\newblock \url{https://github.com/cgleggett/FCS-GPU}.

\end{thebibliography}
\end{document}